\documentclass[aps,prl,superscriptaddress,amsmath,amssymb,twocolumn,nofootinbib]{revtex4}

\usepackage[utf8]{inputenc}
\usepackage{color}
\usepackage{graphicx}
\usepackage{dashbox}
\usepackage{physics}

\def\half{\frac{1}{2}}
\def\kB{k_{\rm B}}

\begin{document}

\thispagestyle{empty}

\title{The effect of GREA on the gravitational collapse of neutron stars}

\author{Juan Garc\'ia-Bellido}\email[]{juan.garciabellido@uam.es}
\affiliation{Instituto de F\'isica Te\'orica UAM-CSIC, Universidad Auton\'oma de Madrid, Cantoblanco 28049 Madrid, Spain}

\date{\today}

\preprint{IFT-UAM/CSIC-23-16}

\begin{abstract}

General Relativistic Entropic Acceleration (GREA) gives a general framework in which to study multiple out-of-equilibrium phenomena in the context of general relativity, like the late accelerated expansion of the universe or the formation of galaxies and the large scale structure of the universe. Here we analyze the conditions for collapse of a star of degenerate neutrons in the presence of entropy production due to the gravitational collapse itself. We find that the final mass and radius of the neutron star differs from that obtained with the adiabatic Tolman-Oppenheimer-Volkoff equation by a factor of order 15\%. We also find that the minimum mass of a neutron star is $\sim 1.1\,M_\odot$ and the maximum mass around $2.4\,M_\odot$. We discuss the possible implications for the search and interpretation of binary coalescing systems like  neutron stars and neutron star-black holes detectable via their multimessenger (gravitational and electromagnetic wave) emission upon merging.

\end{abstract}
\maketitle

\section{I. Introduction}

The physics of compact objects like black holes and neutron stars is a topic of maximum interest for the community, both in astrophysics and fundamental physics~\cite{Shapiro:1983du}. It encounters the most extreme environments where gravity is important and where we can test the theory of general relativity.

The physics of neutron stars (NS) has become an active field of research~\cite{Faber:2012rw} thanks to the recent detection of gravitational waves (GW) from the merger of two neutron stars in GW170817~\cite{LIGOScientific:2017zic}, which has opened the possibility to explore directly the inner structure of neutron stars and their equation of state~\cite{GuerraChaves:2019foa}.

Moreover, the distribution in size and masses of neutron stars has important consequences for the astrophysics of stellar evolution and their collapse through supernovae explosions, as well as in the interpretation of GW events as coming from NSBH binaries or Black Hole (BH) binaries, as in GW190425~\cite{LIGOScientific:2020aai} and GW190814~\cite{LIGOScientific:2020zkf,Clesse:2020ghq}.

On the other hand, our understanding of the nature and stability of stars in general, and neutron stars in particular, rely on specific assumptions about their equations of state and hydrodynamical equilibrium in the context of general relativity.
These boundary conditions assume the coarsegrained macroscopic entropy to be conserved upon gravitational collapse. We explore in this paper the effects that entropic forces associated with gravitational collapse have on the final state of neutron stars, and in particular their masses and radii.

While fundamental interactions can be derived from an action principle among fundamental particles, thermodynamics arises from the coarsegraining of microscopic degrees of freedom in the context of statistical mechanics, and can be incorporated into the action as phenomenological constraints. In a recent work~\cite{Espinosa-Portales:2021cac}, we described a generally covariant formalism for out-of-equilibrium phenomena in the context of gravitational systems, and thus modified the equations of general relativity (GR) to include dissipative phenomena and the growth of entropy. We found that the thermodynamical constraint introduced a new term in the matter energy-momentum tensor which could be understood as a bulk viscosity term. As we will describe in the next section, such a viscous term induces a negative pressure that can account for what we called general relativistic entropic acceleration (GREA), which could explain the present acceleration of the universe from first principles~\cite{Garcia-Bellido:2021idr}, without the need to introduce an {\em ad hoc} and extremely finetuned cosmological constant. Such a scenario is consistent with the present cosmological observations~\cite{Arjona:2021uxs} and could be tested in the near future with precise observations from deep galaxy surveys.

In this paper we explore the consequences of GREA on the gravitational collapse of neutron stars, and we derive a range of allowed masses for neutron stars that differs from those obtained in the absence of entropy production, irrespective of the internal equation of state. Present astrophysical data already suggest that such a range is in better agreement with observations, and in the future the determination of their masses and radii via the detection of gravitational waves from BNS mergers will allow one to constrain better the equation of state in the interior of neutron stars.

\section{II. Gravitational collapse in GREA}

Here we descrive the set of equations that we will resolve in the context of gravitational collapse of degenerate neutron matter when there is entropy production due to gravitational collapse.
The GREA equations are a modification of GR~\cite{Espinosa-Portales:2021cac}
\begin{equation}
	\label{eq:GREA}
	G_{\mu\nu} = R_{\mu\nu} - \half R g_{\mu\nu} = \kappa\, (T_{\mu\nu} - f_{\mu\nu}) \equiv \kappa\,{\cal T}_{\mu\nu}\,,
\end{equation}
where $f_{\mu\nu}$ satisfies the thermodynamic constraint
$$\frac{\partial{\cal L}_m}{\partial S}\delta S = \sqrt{-g}f_{\mu\nu}\delta g^{\mu\nu}\,,$$
arising from the fundamental laws of thermodynamics
\begin{eqnarray}
	\label{eq:TdS}
	-dW = - \vec F\cdot d\vec x &=& dU + \left(P - T\frac{dS}{dV}\right)dV \nonumber \\
 &\equiv& dU + \tilde P\,dV
\end{eqnarray}
where we have defined an {\em effective} pressure $\tilde P$ which reduces to the usual fluid pressure  $P$ in the absence of entropy production. This extra component to the Einstein equations can be interpreted as an effective bulk viscosity term of a real (non-ideal) fluid~\cite{Espinosa-Portales:2021cac}
\begin{equation}
	\label{eq:fmunu}
	f_{\mu\nu} = \zeta\,D_\lambda u^\lambda \,(g_{\mu\nu}+u_\mu u_\nu) = \zeta\,\Theta\,h_{\mu\nu} \,,
\end{equation}
such that the covariantly-conserved energy-momentum tensor becomes
\begin{eqnarray}
	\label{eq:Tmunu}
	{\cal T}^{\mu\nu} &=& P\,g^{\mu\nu} + (\epsilon + P)u^\mu u^\nu -  \zeta\,\Theta\,h^{\mu\nu} \\ &=& \tilde P\,g^{\mu\nu} + (\epsilon + \tilde P)u^\mu u^\nu\,,
\end{eqnarray}
and, imposing the thermodynamic constraint~(\ref{eq:TdS}), the bulk viscosity coefficient $\zeta$ can be written as
\begin{equation}
	\label{eq:zeta}
	\zeta = \frac{T}{\Theta}\frac{dS}{dV}\,.
\end{equation}
In the case of an expanding universe, $\Theta=\frac{d}{dt}\ln V = 3H$ and the coefficient becomes $\zeta = T\dot S/(9H^2a^3)$, see~\cite{Garcia-Bellido:2021idr}, with $S$ the entropy per comoving volume of the Universe. Entropy production therefore implies $\zeta > 0$.

Note that the energy-momentum tensor is still diagonal, ${\cal T}^\mu_{\ \ \nu} = {\rm diag}(-\epsilon,\,\tilde P,\,\tilde P,\,\tilde P)$, and that the $00$ component is unchanged with respect to GR. Only the $ij$ com\-ponent has the entropy-growth dependence via $\tilde P$.

The Raychaudhuri equation for geodesic motion in the absence of shear ($\sigma=0$) and vorticity ($\omega=0$) is
\begin{eqnarray}
	\label{eq:Ray} \nonumber
	\frac{D}{d\tau}\Theta + \frac{1}{3}\Theta^2 &=& - \sigma_{\mu\nu}\sigma^{\mu\nu} + \omega_{\mu\nu}\omega^{\mu\nu} - R_{\mu\nu} u^\mu u^\nu \\ &=& - \kappa\left(T_{\mu\nu}u^\mu u^\nu + \half T^\lambda_{\ \,\lambda} - \frac{3}{2}\zeta\Theta\right) \\ &=& - \frac{\kappa}{2}(\epsilon + 3\tilde P) = -\frac{\kappa}{2}\left(\epsilon + 3P - 3T\frac{dS}{dV}\right) \nonumber\,.
\end{eqnarray}
Due to the extra entropic term in the effective pressure $\tilde P$, even for matter that satisfies the strong energy condition, $\epsilon + 3P > 0$, it is possible to prevent gravitational collapse, i.e. $\dot\Theta + \Theta^2/3 > 0$, as long as the production of entropy is significant enough, $3TdS/dV > (\epsilon + 3P) > 0$.

\section{III. TOV equations in GREA}

The gravitational collapse of a fluid of $N$ relativistic particles of mass $m$ has an associated Boltzmann entropy
\begin{eqnarray}
	\label{eq:Boltzmann} \nonumber
	S &=& \kB\,\ln\int\!\!\int d^3q_1\dots d^3q_N\, d^3p_1\dots d^3p_N\, f(q_i\,p_i) \\ &=& \kB\,N\,\ln\left[\frac{V}{N}\left(\frac{\langle p^2\rangle}{4\pi^2\hbar^2} \right)^{3/2} \right]\,.
\end{eqnarray}
The gravitational collapse of a star occurs far from equilibrium. On the one hand, the collapse reduces phase-space volume due to the spatial contraction. On the other hand, the fluid is heated upon contraction and thus increases its average momentum-squared, $\langle p^2\rangle$. If we use the virial theorem $\langle K\rangle = -\half\langle V\rangle$ for a bounded system under gravity,
\begin{equation}
	\label{eq:Virial}
	\half m\,\langle v^2\rangle = \frac{3}{2}\kB T = \frac{GM(r)m}{2\,r}\,,
\end{equation}
we see that 
\begin{eqnarray}
	\label{eq:Entropy} 
	S &=& \kB\,N\,\ln\left[\frac{V}{N}\left(\frac{3m\,\kB T}{4\pi^2\hbar^2} \right)^{3/2} \right] \\ &=& \kB\,N\,\ln\left[\frac{V}{N}\left(\frac{m^2c^2}{4\pi^2\hbar^2}\cdot\frac{GM(r)}{N c^2r} \right)^{3/2} \right] \,. \nonumber
\end{eqnarray}
and thus the dependence of the entropy on the volume $V$ is always positive, $dS/dV > 0$, and therefore a contraction, $dV<0$, implies a local decrease in entropy, $dS < 0$. The second law of Thermodynamics is ensured by the fact that the heating of the star upon contraction will induce photon and neutrino production (via electro-weak interactions) which will radiate away, producing a flux of entropy out of the star that overall ensures $dS > 0$. In most cases, even matter is shed away in gigantic thermonuclear reactions known as supernovae.

Let us now study the general relativistic equations that determine the final state of collapse that leads to a neutron star (NS) with energy density $\epsilon(\rho) = \rho\,c^2 + \varepsilon$, and pressure $P(\rho)$, such that $T^\mu_{\ \,\nu} = {\rm diag}(-\epsilon,\,P,\,P,\,P)$. Let us assume a metric for the interior of the star given by
\begin{equation}
	\label{eq:metric}
	ds^2 = - e^{2\Phi(r)}dt^2 + e^{2\Psi(r)}dr^2 + r^2d\Omega^2 \,,
\end{equation}
Where $\Phi$ and $\Psi$ are only $r$-dependent. The $00$-component of the Einstein equations is not modified, so we have the solution~\cite{Tolman:1939jz,Oppenheimer:1939ne}
\begin{equation}
	\label{eq:Psi}
	\Psi(r) = - \half \ln \left(1 - \frac{2GM(r)}{r}\right)\,,
\end{equation}
where 
\begin{equation}
	\label{eq:Mr}
	   M(r) = \int 4\pi\,r^2\,dr\,\epsilon(r)\,.
\end{equation}
We will now assume that the spherically symmetric collapse has an associated entropy production (\ref{eq:Entropy}), so that the condition of hydrostatic equilibrium, $u^\mu=((-g_{00})^{-1/2},\,0),\ \partial_0 g_{\mu\nu} = 0$, remains valid, but where the covariant conservation of the {\em full} energy-momentum tensor is now
\begin{equation}
	\label{eq:Tcons}
	   D_\mu{\cal T}^\mu_{\ \nu} = \partial_\nu\tilde P + (\epsilon + \tilde P)\,\partial_\nu\ln\sqrt{-g_{00}} = 0\,,
\end{equation}
while the radial component gives the $\Phi(r)$ equation
\begin{equation}
	\label{eq:DTr}
	   \partial_r(P - \zeta\Theta) + (\epsilon + P - \zeta\Theta)\,\Phi'(r) = 0\,.
\end{equation}
Note that we differentiate here between the total energy density $\epsilon(\rho)$ and the rest-mass density $\rho=m\,n$.

The spatial part of the Einstein equations gives rise to the modified Tolman-Oppenheimer-Volkoff equation
\begin{equation}
	\label{eq:TOV}
	   \tilde P'(r) = - \frac{GM(r)}{r^2}\left(\epsilon(r) + \tilde P(r)\right)\frac{1+4\pi r^3\tilde P(r)/M(r)}{1-2GM(r)/r}\,.
\end{equation}

The neutron fluid may have an equation of state $P(\rho)$ which is undetermined in principle, and may go from fully relativistic to non-relativistic. The individual neutrons have energies $E^2(p)=m^2c^4 + p^2c^2$ inside the star. If we assume the neutron fluid becomes degenerate with Fermi distribution $f_F(p) = \theta(p_F-p)$, with Fermi momentum $p_F$, then the number density is ($g=2$ for two spin states)
\begin{eqnarray}
	\label{eq:nF}\nonumber
	   n_F &=& 2\int\frac{d^3{\bf p}}{\hbar^3(2\pi)^3}\,f_F(p) \\ &=& 2\int_0^{p_F}\!\frac{4\pi p^2dp}{\hbar^3(2\pi)^3} = \frac{p_F^3}{3\pi^2\hbar^3}\,,
\end{eqnarray}
so that
\begin{equation}
	\label{eq:xF}
	   x_F \equiv \frac{p_F}{mc} = \left(\frac{3}{8\pi}\right)^{1/3}\frac{h}{mc}\,n_F^{1/3} = A\,\rho^{1/3}\,,
\end{equation}
in terms of the rest-mass density, $\rho = m\,n_F$,
while the pressure becomes
\begin{eqnarray}
	\label{eq:nF}\nonumber
	   P(x_F) &=& 2\int\frac{d^3{\bf p}}{\hbar^3(2\pi)^3}\,\frac{p^2c^2}{3E(p)}\,f_F(p) \\ &=& 2\int_0^{p_F}\!\frac{4\pi p^2dp}{\hbar^3(2\pi)^3}\,\frac{p^2c^2}{3\sqrt{m^2c^4 + p^2c^2}} \nonumber \\ &=& \rho\,\frac{x_F}{4}\,I(x_F)\,,
\end{eqnarray}
where
\begin{equation}
	\label{eq:Ix}
	   I(x) \equiv \frac{1}{2x^4} \left(3\sinh^{-1}(x)+(2x^3-3x)\sqrt{1+x^2}\right)\,,
\end{equation}
with asymptotics: $I(x) = 4x/5$ in the NR limit, $x_F\ll1$, and $I(x) = 1$ in the ultrarelativistic limit, $x_F\gg1$, which corresponds to equations of state $P(\rho)\propto \rho^{5/3}$ and $\rho^{4/3}$ in the NR and UR limits, respectively. The speed of sound becomes
\begin{equation}
	\label{eq:cs2}
	   c_s^2 = \frac{dP(\rho)}{d\rho} = \frac{1}{3}\frac{x_F^2}{\sqrt{1+x_F^2}}\,.
\end{equation}

We will now take into account the laws of thermodynamics out-of-equilibrium, $d(\epsilon V) = -PdV + TdS$, to obtain a solution to the modified TOV equation (\ref{eq:TOV}), using $V=Nm/\rho$ for fixed number of particles $N$ and Eq.~(\ref{eq:Entropy}), 
\begin{equation}
	\label{eq:TS}
	   d\left(\frac{\epsilon(\rho)}{\rho}\right) = \tilde P(\rho) \,\frac{d\rho}{\rho^2} = \left(P(\rho) - \rho\,\frac{\kB T}{mc^2}\right)\frac{d\rho}{\rho^2}\,,
\end{equation}
which can be integrated
\begin{equation}
	\label{eq:eps}
	   \epsilon(\rho) = \rho\,H(x_F) - \frac{\kB T}{mc^2} \,\rho\,\ln\rho\,,
\end{equation}
with
\begin{equation}
	\label{eq:H}
	   H(x) \equiv \frac{3}{8x^3} \left((2x^3+x)\sqrt{1+x^2} - \sinh^{-1}(x)\right)  \,.
\end{equation}
Taken together, we see that the specific enthalpy is
\begin{equation}
	\label{eq:epsP}
	   \epsilon'(\rho) = \frac{\epsilon(\rho) + \tilde P(\rho)}{\rho} = \sqrt{1+x_F^2} -  \frac{\kB T}{mc^2} (\ln\rho + 1) \,.
\end{equation}
Note that we recover the usual expressions in the limit $\kB T \ll mc^2$, where the entropic force contribution to gravitational collapse is negligible. However, in our case, due to the Virial theorem, gravitational collapse brings together a heating up of the neutron fluid, with an important entropy production, so we must take it into account. As we well see, however, the actual temperatures inside the neutron star never reach more than around 80 MeV, well below the neutron mass, so we are still in the degenerate Fermi gas approximation.

We can solve simultaneously the TOV Eq.~(\ref{eq:TOV}) and the mass equation, $M'(r) = 4\pi r^2 \epsilon(r)$, by choosing as "radial" coordinate the rest-mass density $\rho$,
\begin{equation}
    \label{eq:system}
\begin{array}{rl}
    {\displaystyle  \frac{dM}{d\rho} }
     \!&= {\displaystyle 
     4\pi \,r(\rho)^2\epsilon(\rho)\,\frac{dr}{d\rho} }\\[5mm] 
    {\displaystyle \frac{dr}{d\rho} }
    \!&= {\displaystyle \frac{\tilde P'(\rho)}{\tilde P'(r)} = \frac{P'(\rho) - \kB T/mc^2}{\tilde P'(r)} }
\end{array}
\end{equation}
where in the denominator we substitute the TOV Eq.~(\ref{eq:TOV}) as a function of $\rho$. We can now take the radial dependence of the temperature from the Virial theorem, Eq.~(\ref{eq:Virial}), 
\begin{equation}
	\label{eq:kT}
	   \frac{3\kB T}{mc^2} = \frac{GM(r)}{c^2r}\,.
\end{equation}

The best way to solve the equations is to choose some units and write them as dimensionless equations. We will choose the solar mass $M_\odot$ as unit of mass, the solar Schwarzschild radius $r_S = 2GM_\odot/c^2 = 3$ km as unit of distance and $\rho_0 = 3M_\odot/(4\pi r_S^3) = 1.8\times10^{16}$ g/cm$^3$ as unit of mass density, which is about 90 times the mean density of a NS. In this case, the TOV equations~(\ref{eq:system}) are written in terms of $\bar m = M/M_\odot$, $\bar r=r/r_S$ and $\bar \rho=\rho/\rho_c$.

We solve the coupled equations by varying the central density $\rho_c = \rho(r=0)$ in units of $\rho_0$ until the density drops to zero. At that moment the star ends, with radius $R=r(\rho=0)$, and the exterior metric becomes that of Schwarzschild with $M=M(R)$. 

\begin{figure}
	\centering
	\includegraphics[width=\linewidth]{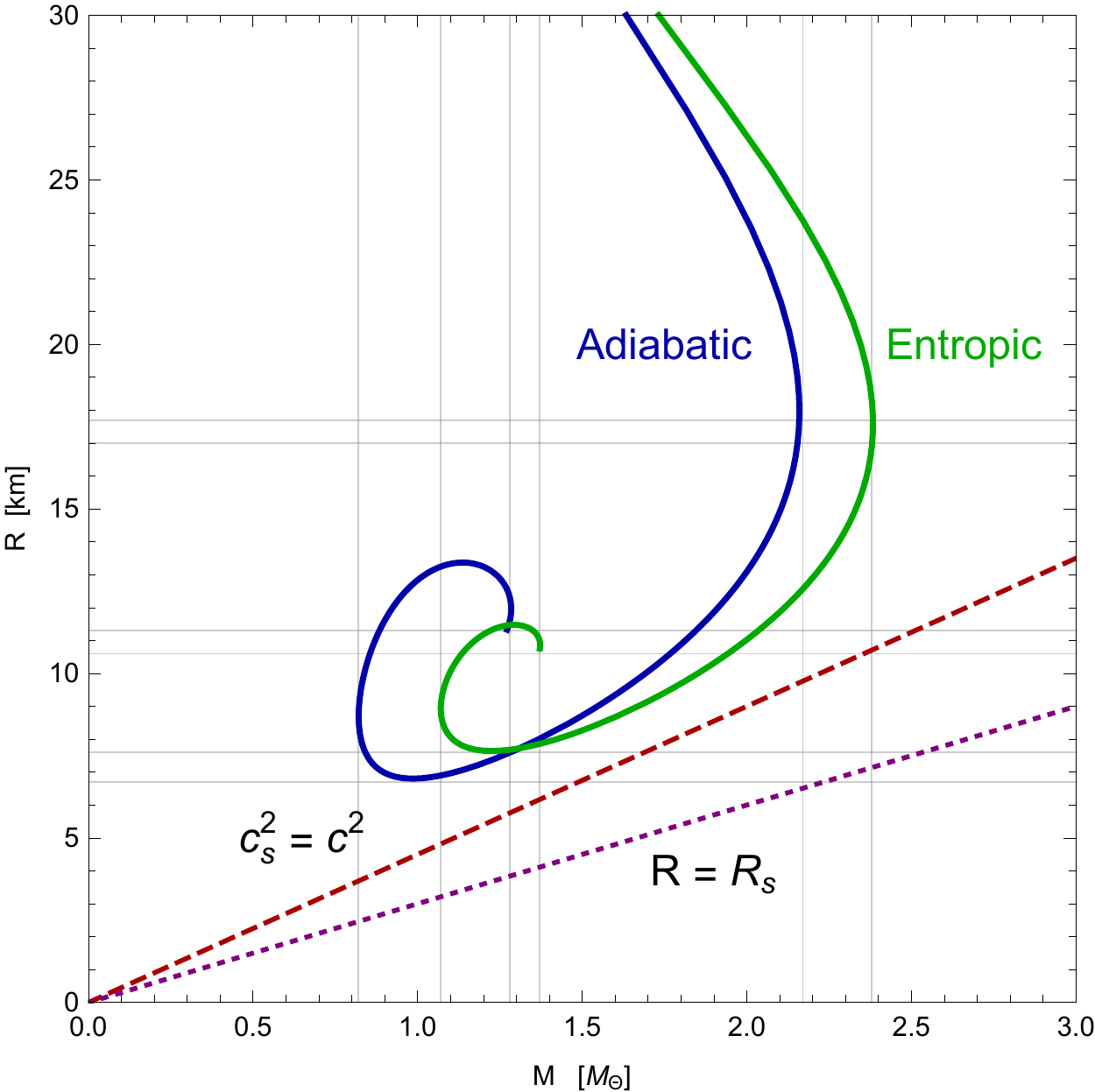}
	\caption{The plane $(M,\,R)$ for neutron stars evolving through out-of-equilibrium gravitational collapse. The blue curve corresponds to the usual TOV-Ad solutions, while the green curve arises from the modified TOV-GREA equations. The red dashed line is the lower limit imposed by the speed of sound, while the purple dotted line bounds the collapse to a black hole. }
	\label{fig:RM}
\end{figure}

In Fig.~\ref{fig:RM} we show the plane $(M,\,R)$ for a series of values of the central density, from $\rho_c=10^{-3}\rho_0$ in the top left, all the way until $\rho_c=100\rho_0$ at the center, for both adiabatic and entropic collapse. Overall the $(M,\,R)$ curves look very similar in the two cases. There is, however, a difference in the minimum and maximum masses attained for neutron stars in hydrostatic equilibrium. For the entropic case, the minimum mass is always above $1.1\,M_\odot$ and the maximum mass always below $2.4\,M_\odot$, which seems to be in better agreement with astrophysical observations than the range for adiabatic case.

\begin{figure}
	\centering
	\includegraphics[width=\linewidth]{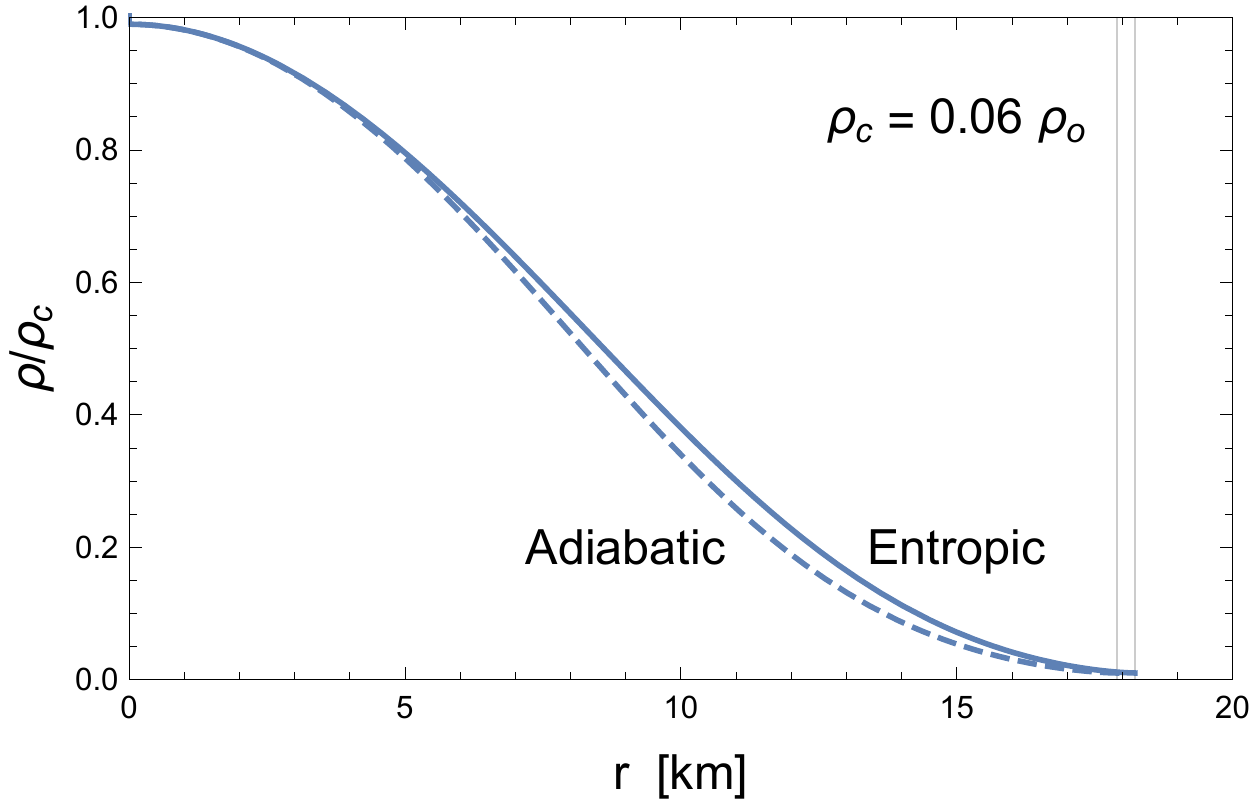}\\[5mm]
	\includegraphics[width=\linewidth]{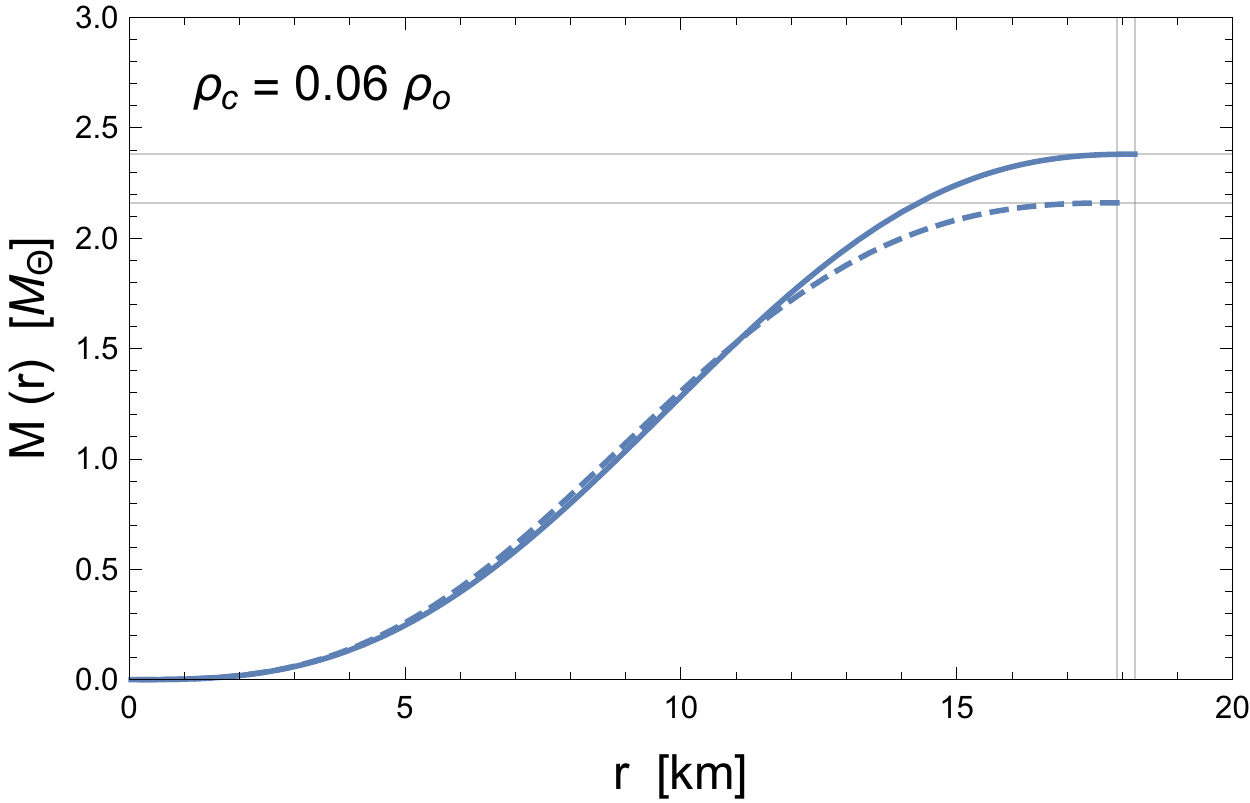}\\[5mm]
	\includegraphics[width=\linewidth]{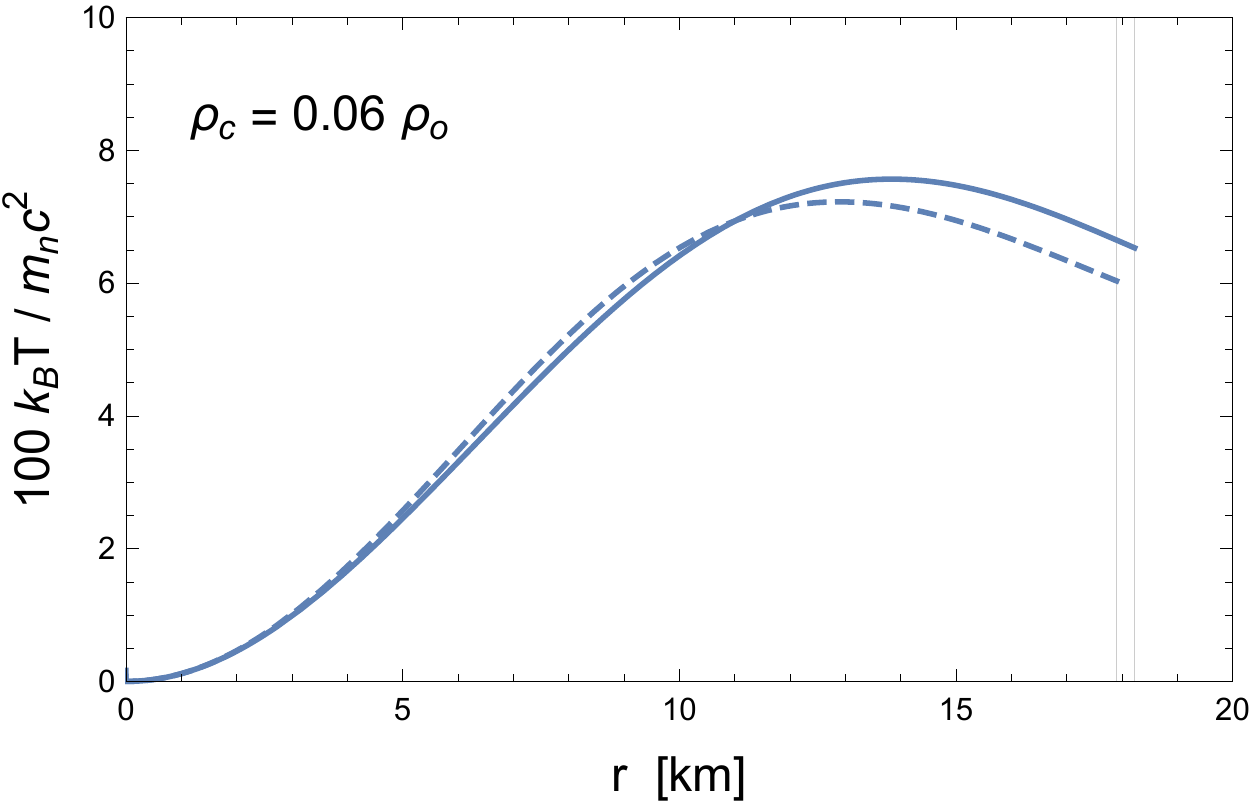}
	\caption{The density, mass and temperature profiles inside the neutron star in hydrostatic equilibrium for adiabatic (dashed line) and entropic (continuous line) collapse. The difference is only appreciable in the outer parts of the star. For these figures, we have chosen the core density $\rho_c=0.06\,\rho_0$ for the NS with the largest possible mass in both cases.}
	\label{fig:RhoMR}
\end{figure}

In Fig.~\ref{fig:RhoMR} we show the radial profiles of the star density (in units of the central core density), mass (in solar masses) and temperature (in units of the neutron mass over 100). It is clear that the entropic forces act mostly in the exterior part of the star, where the temperature is non-negligible. This effect is enough to produce masses of neutron stars that are around 15\% larger than those in the adiabatic solutions.

\section{V. Conclusions}

We have studied the effect of the general relativistic entropic force induced by the entropy production during gravitational collapse of a neutron star on the final state in hydrodynamic equilibrium of degenerate nuclear matter. We solve the modified Tolman-Oppenheimer-Volkoff equations with the extra pressure term proportional to the temperature inside the neutron star and the growth of entropy, by varying the central density and finding the matter profiles inside the star.

We find that the NS with the largest core densities have masses and radii, $(M(M_\odot),\,R({\rm km})) = (1.28,\,10.6)$ and $(1.37,\,11.3)$ for TOV-Ad and TOV-GREA, respectively, which only differ by about 15\%. On the other hand, the maximum and minimium masses are rather different, $(M_{\rm max}(M_\odot),\,R({\rm km})) = (2.17,\,17.7)$ and $(2.38,\,17.1)$ and $(M_{\rm min}(M_\odot),\,R({\rm km})) = (0.82,\,8.9)$ and $(1.07,\,8.9)$  for TOV-Ad and TOV-GREA, respectively. These results suggest that, irrespective of the internal equation of state, the entropic forces associated with gravitational collapse have an important effect on the final equilibrium profiles and the actual masses and radii of neutron stars.

These results may have important consequences for our interpretation of the nature of the compact bodies that form inspiraling binaries, and finally merge emitting a burst of gravitational waves that can be measured with laser interferometers like LIGO-Virgo-KAGRA~\cite{Coupechoux:2023fqq} and, in the future, the Einstein Telescope~\cite{Maggiore:2019uih}. The investigation of the NS equation of state derived from these collisions~\cite{GuerraChaves:2019foa} will thus have to be implemented in the context of the entropic forces acting upon gravitational collapse down to stable configurations that later on form coalescing binaries.

\section{Acknowledgements}

The author acknowledges support from the Research Project PID2021-123012NB-C43 [MICINN-FEDER], and the Centro de Excelencia Severo Ochoa Program CEX2020-001007-S. 

\bibliography{main}

\begin{thebibliography}{14}
\expandafter\ifx\csname natexlab\endcsname\relax\def\natexlab#1{#1}\fi
\expandafter\ifx\csname bibnamefont\endcsname\relax
  \def\bibnamefont#1{#1}\fi
\expandafter\ifx\csname bibfnamefont\endcsname\relax
  \def\bibfnamefont#1{#1}\fi
\expandafter\ifx\csname citenamefont\endcsname\relax
  \def\citenamefont#1{#1}\fi
\expandafter\ifx\csname url\endcsname\relax
  \def\url#1{\texttt{#1}}\fi
\expandafter\ifx\csname urlprefix\endcsname\relax\def\urlprefix{URL }\fi
\providecommand{\bibinfo}[2]{#2}
\providecommand{\eprint}[2][]{\url{#2}}

\bibitem[{\citenamefont{Shapiro and Teukolsky}(1983)}]{Shapiro:1983du}
\bibinfo{author}{\bibfnamefont{S.~L.} \bibnamefont{Shapiro}} \bibnamefont{and}
  \bibinfo{author}{\bibfnamefont{S.~A.} \bibnamefont{Teukolsky}},
  \emph{\bibinfo{title}{{Black holes, white dwarfs, and neutron stars: The
  physics of compact objects}}} (\bibinfo{year}{1983}), ISBN
  \bibinfo{isbn}{978-0-471-87316-7}.

\bibitem[{\citenamefont{Faber and Rasio}(2012)}]{Faber:2012rw}
\bibinfo{author}{\bibfnamefont{J.~A.} \bibnamefont{Faber}} \bibnamefont{and}
  \bibinfo{author}{\bibfnamefont{F.~A.} \bibnamefont{Rasio}},
  \bibinfo{journal}{Living Rev. Rel.} \textbf{\bibinfo{volume}{15}},
  \bibinfo{pages}{8} (\bibinfo{year}{2012}), \eprint{1204.3858}.

\bibitem[{\citenamefont{Abbott et~al.}(2017)}]{LIGOScientific:2017zic}
\bibinfo{author}{\bibfnamefont{B.~P.} \bibnamefont{Abbott}}
  \bibnamefont{et~al.} (\bibinfo{collaboration}{LIGO Scientific, Virgo,
  Fermi-GBM, INTEGRAL}), \bibinfo{journal}{Astrophys. J. Lett.}
  \textbf{\bibinfo{volume}{848}}, \bibinfo{pages}{L13} (\bibinfo{year}{2017}),
  \eprint{1710.05834}.

\bibitem[{\citenamefont{Guerra~Chaves and
  Hinderer}(2019)}]{GuerraChaves:2019foa}
\bibinfo{author}{\bibfnamefont{A.}~\bibnamefont{Guerra~Chaves}}
  \bibnamefont{and} \bibinfo{author}{\bibfnamefont{T.}~\bibnamefont{Hinderer}},
  \bibinfo{journal}{J. Phys. G} \textbf{\bibinfo{volume}{46}},
  \bibinfo{pages}{123002} (\bibinfo{year}{2019}), \eprint{1912.01461}.

\bibitem[{\citenamefont{Abbott
  et~al.}(2020{\natexlab{a}})}]{LIGOScientific:2020aai}
\bibinfo{author}{\bibfnamefont{B.~P.} \bibnamefont{Abbott}}
  \bibnamefont{et~al.} (\bibinfo{collaboration}{LIGO Scientific, Virgo}),
  \bibinfo{journal}{Astrophys. J. Lett.} \textbf{\bibinfo{volume}{892}},
  \bibinfo{pages}{L3} (\bibinfo{year}{2020}{\natexlab{a}}),
  \eprint{2001.01761}.

\bibitem[{\citenamefont{Abbott
  et~al.}(2020{\natexlab{b}})}]{LIGOScientific:2020zkf}
\bibinfo{author}{\bibfnamefont{R.}~\bibnamefont{Abbott}} \bibnamefont{et~al.}
  (\bibinfo{collaboration}{LIGO Scientific, Virgo}),
  \bibinfo{journal}{Astrophys. J. Lett.} \textbf{\bibinfo{volume}{896}},
  \bibinfo{pages}{L44} (\bibinfo{year}{2020}{\natexlab{b}}),
  \eprint{2006.12611}.

\bibitem[{\citenamefont{Clesse and Garc\'ia-Bellido}(2022)}]{Clesse:2020ghq}
\bibinfo{author}{\bibfnamefont{S.}~\bibnamefont{Clesse}} \bibnamefont{and}
  \bibinfo{author}{\bibfnamefont{J.}~\bibnamefont{Garc\'ia-Bellido}},
  \bibinfo{journal}{Phys. Dark Univ.} \textbf{\bibinfo{volume}{38}},
  \bibinfo{pages}{101111} (\bibinfo{year}{2022}), \eprint{2007.06481}.

\bibitem[{\citenamefont{Espinosa-Portales and
  Garc\'ia-Bellido}(2021)}]{Espinosa-Portales:2021cac}
\bibinfo{author}{\bibfnamefont{L.}~\bibnamefont{Espinosa-Portales}}
  \bibnamefont{and}
  \bibinfo{author}{\bibfnamefont{J.}~\bibnamefont{Garc\'ia-Bellido}},
  \bibinfo{journal}{Phys. Dark Univ.} \textbf{\bibinfo{volume}{34}},
  \bibinfo{pages}{100893} (\bibinfo{year}{2021}), \eprint{2106.16012}.

\bibitem[{\citenamefont{Garc\'ia-Bellido and
  Espinosa-Portales}(2021)}]{Garcia-Bellido:2021idr}
\bibinfo{author}{\bibfnamefont{J.}~\bibnamefont{Garc\'ia-Bellido}}
  \bibnamefont{and}
  \bibinfo{author}{\bibfnamefont{L.}~\bibnamefont{Espinosa-Portales}},
  \bibinfo{journal}{Phys. Dark Univ.} \textbf{\bibinfo{volume}{34}},
  \bibinfo{pages}{100892} (\bibinfo{year}{2021}), \eprint{2106.16014}.

\bibitem[{\citenamefont{Arjona et~al.}(2022)\citenamefont{Arjona,
  Espinosa-Portales, Garc\'\i{}a-Bellido, and Nesseris}}]{Arjona:2021uxs}
\bibinfo{author}{\bibfnamefont{R.}~\bibnamefont{Arjona}},
  \bibinfo{author}{\bibfnamefont{L.}~\bibnamefont{Espinosa-Portales}},
  \bibinfo{author}{\bibfnamefont{J.}~\bibnamefont{Garc\'\i{}a-Bellido}},
  \bibnamefont{and} \bibinfo{author}{\bibfnamefont{S.}~\bibnamefont{Nesseris}},
  \bibinfo{journal}{Phys. Dark Univ.} \textbf{\bibinfo{volume}{36}},
  \bibinfo{pages}{101029} (\bibinfo{year}{2022}), \eprint{2111.13083}.

\bibitem[{\citenamefont{Tolman}(1939)}]{Tolman:1939jz}
\bibinfo{author}{\bibfnamefont{R.~C.} \bibnamefont{Tolman}},
  \bibinfo{journal}{Phys. Rev.} \textbf{\bibinfo{volume}{55}},
  \bibinfo{pages}{364} (\bibinfo{year}{1939}).

\bibitem[{\citenamefont{Oppenheimer and Volkoff}(1939)}]{Oppenheimer:1939ne}
\bibinfo{author}{\bibfnamefont{J.~R.} \bibnamefont{Oppenheimer}}
  \bibnamefont{and} \bibinfo{author}{\bibfnamefont{G.~M.}
  \bibnamefont{Volkoff}}, \bibinfo{journal}{Phys. Rev.}
  \textbf{\bibinfo{volume}{55}}, \bibinfo{pages}{374} (\bibinfo{year}{1939}).

\bibitem[{\citenamefont{Coupechoux et~al.}(2023)\citenamefont{Coupechoux,
  Chierici, Hansen, Margueron, Somasundaram, and Sordini}}]{Coupechoux:2023fqq}
\bibinfo{author}{\bibfnamefont{J.~F.} \bibnamefont{Coupechoux}},
  \bibinfo{author}{\bibfnamefont{R.}~\bibnamefont{Chierici}},
  \bibinfo{author}{\bibfnamefont{H.}~\bibnamefont{Hansen}},
  \bibinfo{author}{\bibfnamefont{J.}~\bibnamefont{Margueron}},
  \bibinfo{author}{\bibfnamefont{R.}~\bibnamefont{Somasundaram}},
  \bibnamefont{and} \bibinfo{author}{\bibfnamefont{V.}~\bibnamefont{Sordini}}
  (\bibinfo{year}{2023}), \eprint{2302.04147}.

\bibitem[{\citenamefont{Maggiore et~al.}(2020)}]{Maggiore:2019uih}
\bibinfo{author}{\bibfnamefont{M.}~\bibnamefont{Maggiore}}
  \bibnamefont{et~al.}, \bibinfo{journal}{JCAP} \textbf{\bibinfo{volume}{03}},
  \bibinfo{pages}{050} (\bibinfo{year}{2020}), \eprint{1912.02622}.

\end{thebibliography}

\end{document}